\documentstyle[twocolumn]{jpsj}
\catcode`\@=11
\def\runauthor{Masatoshi Imada}
\def\runtitle{Suppressed Coherence due to Orbital Correlations in the Mn Perovskites}
\def\eqlt{\mathrel{\mathpalette\@vereq<}}  % < over =
\def\eqgt{\mathrel{\mathpalette\@vereq>}}  % > over =
\def\@vereq#1#2{\lower2.5pt\vbox{\baselineskip0pt \lineskip-.5pt
  \ialign{$\m@th#1\hfil##\hfil$\crcr#2\crcr{=}\crcr}}}

\def\simlt{\mathrel{\mathpalette\@vereq<}}  % < over \sim
\def\simgt{\mathrel{\mathpalette\@vereq>}}  % > over \sim
\def\@versim#1#2{\lower2.5pt\vbox{\baselineskip0pt \lineskip-.5pt
  \ialign{$\m@th#1\hfil##\hfil$\crcr#2\crcr{\sim}\crcr}}}

\begin{document}
\renewcommand{\thefootnote}{\fnsymbol{footnote}}
\title{Suppressed Coherence due to Orbital Correlations in the Ferromagnetically Ordered Metallic Phase of Mn Compounds}

\author{Masatoshi Imada}
\inst{Institute for Solid State Physics, University of Tokyo, 
Roppongi, Minato-ku, Tokyo 106, Japan}
\recdate{ \ \ \ \ \ \ \ \ \ \ \ \ \ \ \ \ \ \ \ \    }
\kword
{
LaMnO$_3$, orbital order, orbital correlation, transition metal oxides, incoherent metal, specific heat, metal-insulator transition, Mott transition, scaling theory, quantum critical phenomena, anisotropic hyperscaling, colossal magnetoresistance, double exchange system
}

\abst{
Small Drude weight $D$ together with small specific heat coefficient $\gamma$ observed in the ferromagnetic phase of R$_{1-x}$A$_x$MnO$_3$ (R=La, Pr, Nd, Sm; A=Ca, Sr, Ba) are analyzed in terms of a proximity effect of the Mott insulator.  The scaling theory for the metal-insulator transition with the critical enhancement of orbital correlations toward the staggered ordering of two $e_g$ orbitals such as $3x^2-r^2$ and $3y^2-r^2$ symmetries may lead to the critical exponents of $D \propto \delta^{u}$ and $\gamma \propto \delta^v$ with $u=3/2$ and $v=0$.  The result agrees with the experimental indications.}
%\end{abstract}
\maketitle
\vspace{10mm}
\noindent
Recently metallic phases near the Mott insulator have been studied  systematically due to renewed interest in transition metal compounds such as high-Tc cuprates and manganites.   It has turned out that most of the metallic phases near the Mott insulator show unusually incoherent characteristics with a vanishing or small Drude weight accompanied by broad incoherent tails in the optical conductivity.  This incoherence has been analyzed using the proximity effect of the Mott insulator with the help of the theory of quantum critical scaling for the metal-insulator transition.\cite{Imada1,Imada2}  

In the case of La$_{1-x}$Sr$_x$MnO$_3$ in the metallic region,\cite{Okimoto1,Okimoto2} the optical conductivity has a broad and flat structure which is extended up to eV order,  and has an overall similarity to other transition metal compounds 
when a plausible interband contribution is subtracted. 
This broad incoherent tail extended to large energy values coexists with a sharp true Drude peak which has small weight relative to this incoherent background.  
At $x=0.175$, the Drude weight is only 1/4 or 1/5 of the weight of the incoherent structure.   
In several cases, such as the high-$T_c$ cuprates, the incoherence may be interpreted from a precursor effect of the transition with a critical enhancement of antiferromagnetic spin fluctuations.\cite{Imada1}  
In the case of R$_{1-x}$A$_x$MnO$_3$, however, the metallic phase at small $x$ and low temperatures 
does not have appreciable spin fluctuations.  
Because of the strong Hund's rule coupling, Mn ions in La$_{1-x}$Sr$_x$MnO$_3$ are at a high-spin state with spin-aligned 3$t_{2g}$ and $(1-x)e_g$ electrons per site.  
Although the insulating phase at $x=0$ has the $A$-type antiferromagnetic ordering of this high-spin state, the metallic phase at low temperatures is discontinuously separated from it in terms of spin correlations because of the saturated ferromagnetism caused by the double exchange mechanism. 
Therefore, spin degrees of freedom do not contribute to generate incoherent charge dynamics because of the absence of fluctuations.   

It is known that in the insulating phase at $x=0$, $e_g$ electrons also have a staggered-type ordering of $3x^2-r^2$ and $3y^2-r^2$ orbitals whose short-ranged fluctuations may persist and be critically enhanced in the metallic phase in contrast with the spin degrees of freedom.  
This orbital fluctuation may be the origin of the incoherent charge dynamics.  When the orbital fluctuations remain, a naive expectation would be that the residual entropy arising from the fluctuations may lead to a large specific heat coefficient $\gamma$.  
This is the usual route of the mass enhancement in the Fermi liquid description.\cite{Imada3}  
However, the observed $\gamma \sim 3$-$5$mJ/K$^2$/mole for R$_{1-x}$A$_x$MnO$_3$ at $x \sim 0.2$ indicates a rather small enhancement\cite{Woodfield} (a few times the value for noninteracting  bare electrons and similar to the value expected from the band electron mass) , which is in contrast with the small Drude weight ($\sim 0.012$ times the value expected from $\pi e^2 n/m$ for the electron concentration $n \sim 0.8$ and the bare electron mass $m$) observed at $x=0.175$.\cite{Okimoto1,Okimoto2}  
The spectral weight of the optical conductivity at $x=0.175$ integrated even up to the charge gap value $\Delta_c$ at $x=0$ is yet around 0.06-0.1 of the noninteracting value when we subtract the plausible interband contribution.  This integrated weight includes the Drude weight as well as the incoherent background.  Despite the ambiguity of the cutoff frequency in the definition, this integrated weight is a useful measure for the intraband contribution of the kinetic energy because the averaged total kinetic energy $\langle K\rangle$ is given from the optical conductivity $\sigma(\omega)$ by $\langle K\rangle = \int_{-\infty}^{\infty} \sigma(\omega)d\omega$.    
Since the intraband contribution in the kinetic energy is given by $\pi e^2 n/m$ in the band description, we need to attribute this reduction ($\sim 0.06$-0.1) to the optical mass enhancement of the order around 10-15 as compared to the noninteracting case because the carrier concentration must be close to 1 to account for a large Fermi volume near the Mott insulator.  It should be noted that, in this picture, it is rather difficult to explain the fact that $\gamma$ is small unless we assume a strong wavenumber dependence of the single-particle selfenergy.  
However, it is natural to expect that this kinetic energy is roughly scaled by the doping concentration $x$ for the case with strongly repulsive on-site Coulomb repulsion, because the motion of each hole contributes to the kinetic energy while the sites with singly occupied $e_g$ electrons without holes do not contribute to it if the double occupation of electrons on the same site is suppressed.  
Therefore, the weight may basically be proportional to the hole concentration and hence be scaled by $x$.  
A more serious problem is that this suppressed kinetic energy is exhausted not in the true Drude peak but mostly in the incoherent broad background below $\Delta_c$.   The real Drude weight is even a small fraction ($\sim \frac{1}{4}$-$\frac{1}{8}$) of it as is mentioned above.  In a simple-minded Drude theory, the Drude weight itself is given by $\pi e^2 n/m$, which leads to an even larger mass enhancement ($\sim 80$) which is in sharp contrast with the observed $\gamma$. 

In this paper, I point out that this seemingly puzzling feature is well understood using the scaling theory of metal-insulator transitions under the proximity effect of the Mott insulator if we take into account  spatial anisotropy of orbital fluctuations as expected in Mn compounds.  
I show the possibility that the anisotropic hyperscaling in three dimensions leads to the exponents  $D \propto \delta^{u}$ and $\gamma \propto \delta^v$ with $u=3/2$ and $v=0$, which is consistent with the experimental observations.  In this paper, the implications of this interpretation are discussed.  Several predictions derived from this theory and to be tested experimentally are also presented.  
Recently, the broad incoherent structure in optical conductivity was discussed by Shiba {\it et al.} \cite{Shiba} in terms of interband transitions between the two $e_g$ bands.  
The flat structure in this three-dimensional compound was also argued by assuming the orbital-charge separation in analogy with the spin-charge separation.\cite{Ishihara} 
My approach is from the critical region of the metal-insulator transition.  
This is appropriate for understanding the above mentioned strong correlation effects with the suppressed coherence and is complementary to the weak coupling approach by Shiba {\it et al.} \cite{Shiba} while my approach is in contrast with that from the orbital-charge separation.    

The orbital ordering of LaMnO$_3$ is characterized by an alternating occupation of $3x^2-r^2$ and $3y^2-r^2$ orbitals in the $ab$ plane, which is termed the  ``antiferroorbital" order while the ordering in the $c$ direction is uniform (``ferroorbital" ordering in this direction).  When the doping concentration $x$ is increased and a ferromagnetic metal is realized, no experimental indication of the long-ranged orbital ordering exists.  This suggests that the carrier dynamics destroys the orbital order.  This is similar to the case of the antiferromagnetic order in mother materials of the high-$T_c$ cuprates which is destroyed upon metallization.  
In Mn compounds, within the $ab$ plane, the disappearance of the orbital order is in contrast with the spin degrees of freedom where the ferromagnetic alignment of spins in the $ab$ plane does not change between metals and the Mott insulators and has no effect near the metal-insulator transition.  While the long-ranged orbital order is lost in metals,  reduction of the doping concentration $x$ may lead to a  growth of the short-ranged orbital correlation with a critical enhancement of the fluctuation toward $x=0$.   Although this enhancement  is interrupted by the appearance of the insulating phase at a finite value of $x=x_c$ due to the unavoidable presence of disorder ($x_c \sim 0.15$ for La$_{1-x}$Sr$_x$MnO$_3$), I discuss below the inherent critical region of orbital correlations is extended into the metallic phase.    

When I concentrate on the $ab$ plane, the noninteracting system with a saturated ferromagnetic order has anisotropic shape of the Fermi surface for two $e_g$ bands, because the transfer of the $e_g$ electrons has spatial anisotropy due to the anisotropy of the wavefunction.  The two $e_g$ bands have different anisotropies and hence the Fermi surfaces have different shapes.  On the other hand, in the Mott insulating phase of the interacting system, the Fermi level at $x=0$ has to satisfy the perfect nesting condition because the Fermi level of $3x^2-r^2$ and $3y^2-r^2$ bands must be located in the gap generated by the commensurate periodicity of the orbital order at the wavenumber $(\pi, \pi,0)$.  
This is again a similar situation to the case of the antiferromagnetic order in the cuprates if we use an analogy between spins and orbitals.  
From the consideration of these two limits,  it turns out that the  continuous growth of orbital correlations toward the long-ranged order at $x=0$ has to renormalize and modify the shapes of the Fermi surface for two bands to the same shape of the perfectly nested one in 2D 
 ($ab$ plane) as in the case of the antiferromagnetic order in the cuprates \cite{Miyake}.  Therefore, the original difference and the anisotropy in two bands become progressively irrelevant and near the critical point, the same universality with the case of the usual single-band Hubbard model with spin degeneracy is realized if we consider only a single $ab$ plane for degenerate two bands of spin-polarized electrons.  
In the $ab$ plane, this leads to the same dynamical exponent $z=4$ for the metal-insulator transition as that in the 2D Hubbard model established in the combined analyses of the scaling theory and the quantum Monte Carlo results.\cite{Imada1,AssaadImada,FurukawaImada}  

In contrast to the case of the cuprates, the manganite has a large dispersion in the $c$ direction which makes the universality class of the metal-insulator transition different. Since the orbital correlation along the $c$-axis grows toward uniform order (``ferroorbital" order), the dynamics in the $c$ direction may be normal because the bands are completely filled in this direction for the Mott insulating phase.  
Thus, we do not have any reason to exclude the charge dynamics along the $c$-axis with the $k^2$ dispersion, which leads to the dynamical exponent $z=2$ in the $c$ direction.  
The only constraint from the experimental results may be that the  dispersion (bandwidth) in the $c$-axis direction must be smaller than the Mott gap to guarantee the appearance of the Mott insulating state.  Otherwise, the Mott insulating phase would collapse to a metal.   

In the orbital ordered state, the rotational symmetry among the $a,b$ and $c$ directions are broken with anisotropy between the $ab$ plane and the $c$ direction while the three axes  are symmetric in the absence of the orbital order.  In other words, the directional symmetry breaking may also be absent in addition to the absence of ``antiferroorbital" order of the $3x^2-r^2$ and $3y^2-r^2$ orbitals.  However, the correlation for the directional order may have a correlation length independent of the above ``antiferroorbital" correlation length and indeed the directional correlation length may be longer.  
The hole dynamics severely destroys the antiferroorbital order while the directional order is not necessarily destroyed critically.  
In this case the critical exponents are associated with the ``antiferroorbital" correlation.  

Because of the anisotropic dispersion of charge excitations characterized by
\begin{equation}
\varepsilon(k) = \alpha_x(k_x^2 + k_y^2)^{\frac{p}{2}} + \alpha_z k^q_z, 
\label{1}
\end{equation}
the scaling properties follow anisotropic hyperscaling.  
We have taken the $x,y$ and $z$ directions along the $a,b$ and $c$ axes above, respectively.  
Here, the dynamical exponent in the plane of the antiferroorbital correlation is inferred to be $z=4$, which leads to $p=4$ while the normal dispersion in the direction perpendicular to the plane leads to $q=2$.   
We note that $z$ connects the frequency scale with the inverse of the characteristic length scale by definition and thus is clearly determined from the dispersion.\cite{Imada1}
From this, the characteristic length scales in the plane and perpendicular to the plane defined by $\xi$ and $\xi_z$, respectively, are related with the characteristic energy scale $E_F$ as
\begin{equation}
E_F = \alpha_x\xi^{-p} = \alpha_z\xi_z^{-q},
\label{2}
\end{equation}
with $\alpha_x$ and $\alpha_z$ being constants.  
This gives the scaling for the singular part of the total energy $E_s$, and the doping concentration $\delta$ as
\begin{eqnarray}
\delta & = & \int_0^{\xi^{-1}} {\rm d}k_x {\rm d}k_y \int_0^{\xi_z^{-1}}{\rm d}k_z \nonumber \\
&= & AE_F^{1/\lambda},  \label{2.2}\\
E_s & = & \int_0^{\xi^{-1}}{\rm d}k_x {\rm d}k_y \int_0^{\xi_z^{-1}}{\rm d}k_z \varepsilon(k) \nonumber \\
& = & 2A \left(\frac{1}{p+2} + \frac{1}{2(q+1)}\right) \left(\frac{\delta}{A}\right)^{\lambda +1},
\label{3}
\end{eqnarray}
with 
\begin{equation}
A=\pi(1/\alpha_x)^{2/p}(1/\alpha_z)^{1/q}, \label{4}
\end{equation}
and 
\begin{equation}
\lambda=\left(\frac{2}{p} + \frac{1}{q}\right)^{-1}.  \label{5}
\end{equation}

The density of states is given by 
\begin{equation}
\frac{\partial \delta}{\partial E_F} = {\frac{A}{\lambda}}^{\lambda} \delta^{1-\lambda}.
\label{6}
\end{equation}
In the conventional metal-insulator transitions with $p=q=2$, the density of states is scaled as 
\begin{equation}
\frac{\partial \delta}{\partial E_F} \propto \delta^{1/3},
\label{7}
\end{equation}
while in the case of $p=4$, $q=2$, we obtain 
\begin{equation}
\frac{\partial \delta}{\partial E_F} \propto \delta^0. \label{8}
\end{equation}
In the case of the noninteracting system, aside from the van-Hove singularity, the density of states is of course kept finite with the absence of the transition to the insulating phase.  As compared to this noninteracting case, eq. (\ref{8}) shows the absence of enhancement.  The specific heat coefficient $\gamma$ basicaly follows the same scaling as that of the density of states and we expect 
\begin{equation}
\gamma \propto \delta^{1-\lambda} \label{9}.
\end{equation}

The Drude weight should be anisotropic if the directional order is present.  However, in the absence of the directional order, the Drude weight becomes isotropic, controlled by the direction which has the most incoherent dynamics.  This bottleneck is the $xy$ plane contribution in our case.  The scaling dimension of the Drude weight is determined from
\begin{equation}
D \sim E_s\xi^2, \label{10}
\end{equation}
because the Drude weight is given by the stiffness constant of the system when one imposes a twisted boundary condition in the spatial direction for a finite-size system with the linear dimension $L$ and takes the limit $L\rightarrow\infty$ afterwards.  When one imposes the twist of the phase $\tilde\phi_L$ between $x=0$ and $x=L$, the Drude weight $D$ is given from the vector potential $A_x=c\tilde\phi_L/Le$ and the singular part of the energy density $E_s/L^d$ as 
\begin{equation}
D=2\pi e^2 \frac{\partial^2 E_s}{\partial\tilde\phi_L^2}L^{2-d}. \label{11}
\end{equation}
Using the finite-size scaling $E_s=E_s^{(0)}{\cal E}(\xi/L)L^d$, from the scale invariance,  we obtain eq.(\ref{10}).  Therefore, for $p\geq 2$ and $q=2$, from eqs.(\ref{3}) and (\ref{10}), the Drude weight is scaled as
\begin{equation}
D\propto\delta^{\lambda+1-\frac{2}{p}\lambda}.  \label{12}
\end{equation}
This gives $D\propto\delta$ for the conventional band-insulator-metal transition with $p=q=2$ while we obtain
\begin{equation}
D\propto\delta^{3/2}, \label{13}
\end{equation}
for $p=4$ and $q=2$.  We note that the noninteracting system has the carrier density $n=1-\delta$ and the Drude weight should be scaled not as $D\propto\delta$ but as $D\propto n$ near $\delta=0$.  Therefore, in contrast to the density of states, $D$ shows a severe suppression in the order $\delta^{3/2}$ as compared to the noninteracting system with a large Fermi volume.  

The above scaling form explains why the Drude weight and the specific heat $\gamma$ are both small.  Since the suppression of the Drude weight as compared to a simple-minded expectation from the Fermi liquid is scaled by $\delta^{3/2}$, the suppression may be $\sim 1/14$ at $\delta=0.175$ which may account for the experimental observation $\sim1/80$ if we take into account a difference between the band mass and the bare mass.  In fact, $\gamma$ value is consistent if the band mass is a few times larger than the bare mass.  

The scaling theory predicts specific $\delta=x$ dependence for various physical quantities such as $D\propto \delta^{3/2}$ and $\gamma\propto\delta^0$.  To test the validity of this scaling, it is desirable to analyze $x$ dependence more systematically in additional experiments.  
The second prediction from the present scaling theory is that the shape of the Fermi surface has to be renormalized to a perfectly nested one in the $ab$ plane,  especially at $k_z=0$.  The third prediction is that the orbital correlation is critically enhanced toward $x=0$ with a specific correlation length exponent $\nu = 1/4$ in the $ab$ plane, although the carrier localization at $x<0.15$ interrupts this criticality.  The fourth point which deserves experimental analyses is related to the possible realization of the hyperscaling with $z=4$ in 2D.  
The flat dispersion of $\varepsilon\sim k^4$ is speculated to be realized around $(\pi,0)$ and $(0,\pi)$ in 2D systems.\cite{Imada2}  
Therefore we also expect that this flat dispersion is realized around $(\pi,0,0)$ and $(0,\pi,0)$ in Mn compounds although the absence of the directional order may make the direct observation of this flat structure in metals rather difficult.  

To realize this anisotropic hyperscaling, we have to take into account a strong wavenumber dependence of the selfenergy in the single-particle analysis.  This wavenumber dependence has to reproduce both the renormalization of the Fermi surface to the nested shape and the flatness of the dispersion around $(\pi,0,0)$ in the plane direction.  
This dispersion around $(\pi,0,0)$ may have the form of 
\begin{equation}
\varepsilon(k) = \alpha_x(k_x^4 - k_y^4) + \alpha_zk_z^2, \label{varep}
\end{equation}
as the wavenumber $k$ is measured from the original $(\pi,0,0)$ point.  
It is clear that we need a singular $k$-dependence of the self energy around this point.  
The realization of this singular dependence from a microscopic and analytic calculation deserves further study.  

In summary, a scaling theory for the metal-insulator transition of Mn perovskite compounds is proposed.  The anisotropic hyperscaling well accounts for the observed tiny Drude weight and the small specific heat coefficient $\gamma$.  The proximity effect of the Mott insulator is important under the critical growth of the orbital correlations.  The specific heat coefficient $\gamma$ and the Drude weight D are scaled as $\gamma \propto\delta^0$ and ${\rm D}\propto \delta^{3/2}$ for the doping concentration $\delta$.  

\section*{Acknowledgements}
The author thanks Y. Tokura for the useful discussions on the experimental results.  This work is supported by a Grant-in-Aid for ``Research for the Future" Program from the Japan Society for the Promotion of Science under the project JSPS-RFTF97P01103.  This work is also partially supported by a Grant-in-Aid on Priority Areas ``Anomalous Metallic States near the Mott Transition" from the Ministry of Education, Science, Sports and Culture.

\end{document}